# Red Giant Bound on the Axion-Electron Coupling Revisited

Georg Raffelt

Max-Planck-Institut für Physik

Föhringer Ring 6, 80805 München, Germany

and

Achim Weiss

Max-Planck-Institut für Astrophysik

85748 Garching, Germany

October 3, 1994

**Abstract**

If axions or other low-mass pseudoscalars couple to electrons ("fine structure constant" $\alpha_a$) they are emitted from red giant stars by the Compton process $\gamma + e \to e + a$ and by bremsstrahlung $e + (Z, A) \to (Z, A) + e + a$. We construct a simple analytic expression for the energy-loss rate for all conditions relevant for a red giant and include axion losses in evolutionary calculations from the main sequence to the helium flash. We find that $\alpha_a \lesssim 0.5 \times 10^{-26}$ or $m_a \lesssim 9\,\mathrm{meV}/\cos^2\beta$ lest the red giant core at helium ignition exceed its standard mass by more than $0.025\,\mathcal{M}_\odot$, in conflict with observational evidence. Our bound is the most restrictive limit on $\alpha_a$, but it does not exclude the possibility that axion emission contributes significantly to the cooling of ZZ Ceti stars such as G117–B15A for which the period decrease was recently measured.

# 1 Introduction

The cooling rate of the ZZ Ceti star G117–B15A as determined from the decrease of its pulsation period appears to be somewhat faster than can be accounted for by standard photon cooling. Isern, Hernandez and Garcia-Berro [1] speculated that this discrepancy was an indication for a novel cooling agent, notably for the emission of "invisible axions".

Axions [2] are low-mass pseudoscalar particles that couple to electrons by virtue of a Lagrangian density

$$\mathcal{L}_{\text{int}} = -ig\,\overline{\psi}_e \gamma_5 \psi_e\, a \tag{1}$$

where $g$ is a dimensionless coupling constant, $\psi_e$ is the electron Dirac field, and $a$ the axion field. We shall also use the "axionic fine structure constant"

$$\alpha_a \equiv g^2/4\pi \quad \text{and} \quad \alpha_{26} = \alpha_a/10^{-26}. \tag{2}$$

In a certain class of models (DFSZ axions) the Yukawa coupling is

$$g = 2.8 \times 10^{-14}\, m_{\text{meV}} \cos^2 \beta \tag{3}$$

where $\cos^2 \beta$ is a model-dependent parameter which we shall always set equal to unity, and $m_{\text{meV}}$ is the axion mass $m_a$ in units of $1\,\text{meV} = 10^{-3}\,\text{eV}$. Then, $\alpha_{26} = 0.64 \times 10^{-2}\, m_{\text{meV}}^2$.

The main energy-loss mechanism in a white dwarf is bremsstrahlung emission $e + (Z,A) \to (Z,A) + e + a$. Isern, Hernandez and Garcia-Berro [1] favored an axion mass of $8.4\,\text{meV}$, equivalent to $\alpha_{26} = 0.45$, in order to explain the cooling rate of G117–B15A.

Of course, this interpretation is very speculative and so, naturally one wants to know if it is consistent with other astrophysical phenomena that might be affected by axion emission. For example, the overall white dwarf luminosity function leads to a constraint of $\alpha_{26} \lesssim 1.0$ [3].

Another constraint was derived by Wang [4] who required that axion cooling would not prevent carbon ignition in accreting white dwarfs so that type I supernova explosions can occur. Wang's bound, based on a simple analytic estimate, is $\alpha_{26} \lesssim 6$ or $m_a \lesssim 30\,\text{meV}$.

Horizontal-branch stars have a nondegenerate, helium-burning core which would emit axions dominantly by the Compton process $\gamma + e \to e + a$. A crude bound is based on the requirement that the energy-loss rate should not exceed $100 \, \text{erg} \, \text{g}^{-1} \, \text{s}^{-1}$ or else the HB lifetime would be reduced by more than about a factor of two, in conflict with the observed number of HB stars in globular clusters. Then one finds the bound $\alpha_{26} \lesssim 5$ [5].

The potentially most restrictive argument discussed in the literature was put forth by Dearborn, Schramm and Steigman [6]. They considered the impact of axion emission on red giants near the helium flash; for $\alpha_{26} \gtrsim 0.16$ they found helium ignition to be suppressed entirely which would clearly contradict the mere existence of the horizontal and asymptotic giant branches observed in stellar systems. Unfortunately, they used emission rates which did not take degeneracy effects properly into account; near the center of a red giant before helium ignition they overestimate the energy-loss rate by as much as a factor of 10 (see below). Still, their adjusted limit on $\alpha_{26}$ is only a factor of 2 or 3 above the value favored to explain the cooling rate of G117–B15A, and so, it seems worthwhile to revisit the helium ignition argument with a more appropriate energy-loss rate.

## 2 Energy-Loss Rate

### 2.1 Compton Process

The simplest possibility to produce axions by virtue of their coupling to electrons is the Compton process $\gamma + e \to e + a$ [7]. In the nonrelativistic limit one finds a cross section $\sigma = 4\pi\alpha\alpha_a\omega^2/3m_e^4$ with $\alpha = 1/137$ and $\omega$ the photon energy. A simple integral over the initial-state photon phase space then yields the energy-loss rate per unit mass

$$\epsilon = \frac{160 \, \zeta_6 \, \alpha\alpha_a}{\pi} \frac{Y_e T^6}{m_N m_e^4} F = \alpha_{26} \times 33 \, \text{erg} \, \text{g}^{-1} \, \text{s}^{-1} \, Y_e \, T_8^6 \, F \qquad (4)$$

where $\zeta_6 \approx 1.0173$, $Y_e$ is the number of electrons per baryon, $m_N$ is the nucleon mass which is used for an approximate conversion between the number density of baryons and the mass density of the medium, and $T_8 = T/10^8 \, \text{K}$.

The factor $F$ accounts for relativistic corrections as well as for degeneracy effects and the nontrivial photon dispersion in a medium. For our purposes, the most severe deviation from $F = 1$ occurs at the center of a red giant before the helium flash. Taking $\rho = 10^6 \, \text{g/cm}^3$ and $T = 10^8 \, \text{K} = 8.6 \, \text{keV}$ as nominal values, the plasma frequency is $18 \, \text{keV}$ and the electron Fermi momentum is 409 keV whence the degeneracy parameter is $\eta = (\mu - m_e)/T = 16.7$. Typical blackbody photons have an energy of about $3T$ whence corrections from a "photon mass" remain moderate. Also, relativistic corrections to the emission rate are only about a 30% effect (Fukugita, Watamura and Yoshimura [8]).

These authors also gave a table for $F$ on a grid of $T$ and $\rho$. For a fixed temperature, their values for $F$ slightly *increase* with increasing density, contrary to the expectation that degeneracy effects should *decrease* the emission rate. Upon closer scrutiny we are unable to find a Pauli-blocking factor in their expressions of the phase-space integrals. We believe that the Compton process must be suppressed by electron degeneracy which implies that bremsstrahlung dominates (see below). Therefore, a precise calculation for the degenerate regime is not warranted. In order to interpolate between degenerate and nondegenerate conditions, however, an estimate of the suppression factor $F_{\text{deg}}$ is useful.

In the nonrelativistic limit electron recoils can be neglected so that the initial- and final-states have the same momentum. Therefore, $F_{\text{deg}}$ is the Pauli blocking factor, averaged over all electrons,

$$F_{\text{deg}} = \frac{1}{n_e} \int \frac{2 \, d^3\mathbf{p}}{(2\pi)^3} \frac{1}{e^{(E-\mu)/T} + 1} \left(1 - \frac{1}{e^{(E-\mu)/T} + 1}\right), \tag{5}$$

where $\mu$ is the electron chemical potential and $n_e$ the electron density. Then,

$$F_{\text{deg}} = \frac{1}{n_e \pi^2} \int_{m_e}^{\infty} p \, E \, dE \, \frac{e^x}{(e^x + 1)^2}, \tag{6}$$

where $x \equiv (E - \mu)/T$. For degenerate conditions the integrand is strongly peaked near $x = 0$ (the Fermi surface) so that one may replace $p$ and $E$ with $p_{\text{F}}$ and $E_{\text{F}}$, respectively, and one may extend the lower limit of integration to $-\infty$. The integral then yields $T$ so that

$$F_{\text{deg}} = 3 E_{\text{F}} T / p_{\text{F}}^2, \tag{7}$$

where $n_e = p_{\text{F}}^3/3\pi^2$ was used. A Fermi momentum $p_{\text{F}} = 409$ keV implies $E_{\text{F}} = 655$ keV; with $T = 8.6$ keV this gives $F_{\text{deg}} = 0.10$. Of course, there are relativistic corrections to this result.

## 2.2 Nondegenerate Bremsstrahlung

The nondegenerate bremsstrahlung rate $e + (Z, A) \to (Z, A) + e + a$ was first calculated by Krauss, Moody and Wilczek [9] and $e + e \to e + e + a$ was added by Raffelt [10]. Ignoring screening effects which are a small correction for nondegenerate conditions, and allowing a chemical composition of only hydrogen (mass fraction $X$) and helium (mass fraction $1 - X$) the energy-loss rate is

$$\begin{aligned}\epsilon &= \frac{64}{45}\left(\frac{2}{\pi}\right)^{1/2} \alpha^2 \alpha_a \frac{\rho\, T^{5/2}}{m_N^2 m_e^{7/2}} \left[(1+X) + \frac{(1+X)^2}{2\sqrt{2}}\right] \\ &= \alpha_{26} \times 297\,\mathrm{erg\,g^{-1}\,s^{-1}}\, T_8^{2.5}\, \rho_6 \left[(1+X) + \frac{(1+X)^2}{2\sqrt{2}}\right]\end{aligned} \qquad (8)$$

where $T_8 = T/10^8\,\mathrm{K}$ as before and $\rho_6 = \rho/10^6\,\mathrm{g\,cm^{-3}}$.

## 2.3 Degenerate Bremsstrahlung

The degenerate bremsstrahlung rate was calculated in order to derive a bound on $\alpha_a$ from white dwarf cooling times [3]. In this case screening effects must be included; otherwise the emission rate diverges. As a screening scale the electron Thomas-Fermi wave number was used, a common but incorrect practice, which leads to an underestimate of the screening suppression because the main contribution is from the ions. Of course, because the screening scale enters logarithmically the error remains moderate—a factor of 2 or 3 for the white dwarf cooling rate.

The axion emission rate for very degenerate matter relevant for white dwarfs and the crust of neutron stars was also calculated [11]. The main point was to include ion correlations in a strongly coupled plasma, a condition quantified by the parameter

$$\Gamma = \frac{Z^2\, 4\pi\alpha}{aT} = 0.2275\, \frac{Z^2}{T_8}\left(\frac{\rho_6}{A}\right)^{1/3} \qquad (9)$$

where $Z$ is the charge of the ions, $A$ their atomic mass, and $n$ their density which determines the ion-sphere radius $a = (3/4\pi n)^{1/3}$. For $\Gamma > 178$ the ions arrange themselves in a bcc lattice while for $\Gamma \lesssim 1$ their correlations are weak. In a white dwarf $\Gamma$ is typically between 20 and 150.

However, red giants near helium ignition are hot; for our standard set of parameters we find $\Gamma = 0.57$ which implies that Debye screening is still a reasonable description of the ion correlations. The electrons contribute little to screening because the Thomas-Fermi wave number is much smaller than the Debye scale; otherwise the plasma would not be degenerate. (For our standard red giant conditions the Thomas-Fermi wave number is about 50 keV while the Debye scale for the ions is 222 keV.)

With these approximations one finds for the energy-loss rate [12]

$$\epsilon = \frac{\pi^2 \alpha^2 \alpha_a}{15} \frac{Z^2}{A} \frac{T^4}{m_N m_e^2} F = \alpha_{26} \times 10.8 \, \text{erg g}^{-1} \, \text{s}^{-1} \frac{Z^2}{A} T_8^4 \, F, \tag{10}$$

where

$$F = \frac{2}{3} \log\left(\frac{2+\kappa^2}{\kappa^2}\right) + \left[\frac{2+5\kappa^2}{15} \log\left(\frac{2+\kappa^2}{\kappa^2}\right) - \frac{2}{3}\right] \beta_F^2 + \mathcal{O}(\beta_F^4) \tag{11}$$

with $\beta_F = p_F/E_F$ the velocity at the Fermi surface. With $k_D$ the Debye screening scale of the ions (density $n$, charge $Ze$)

$$\kappa^2 = \frac{k_D^2}{2p_F^2} = \frac{4\pi\alpha \, Z^2 \, n}{T} \frac{1}{2p_F^2}. \tag{12}$$

For helium this is $\kappa^2 = 0.147 \, \rho_6^{1/3}/T_8$. For our benchmark conditions we have $\beta_F^2 = 0.39$ and then $F = 1.7$.

## 2.4 Interpolation Formula

The main region of interest to us is the degenerate red giant core. However, the hydrogen burning shell is entirely nondegenerate and also at a temperature of about $10^8$ K so that a consistent treatment requires to implement axion emission everywhere in the star. To this end we interpolate between the degenerate and nondegenerate bremsstrahlung rates by

$$\epsilon = \left(\epsilon_{\text{ND}}^{-1} + \epsilon_{\text{D}}^{-1}\right)^{-1}. \tag{13}$$

The nondegenerate Compton rate is switched off in the degenerate regime by means of a factor $(1 + F_{\text{deg}}^{-2})^{-1/2}$ where $F_{\text{deg}}$ was given in Eq. (7). In Fig. 1 we show the different rates as well as our interpolation as a function of $\rho$ for $T = 10^8$ K. Interestingly, the total rate is nearly independent of density;

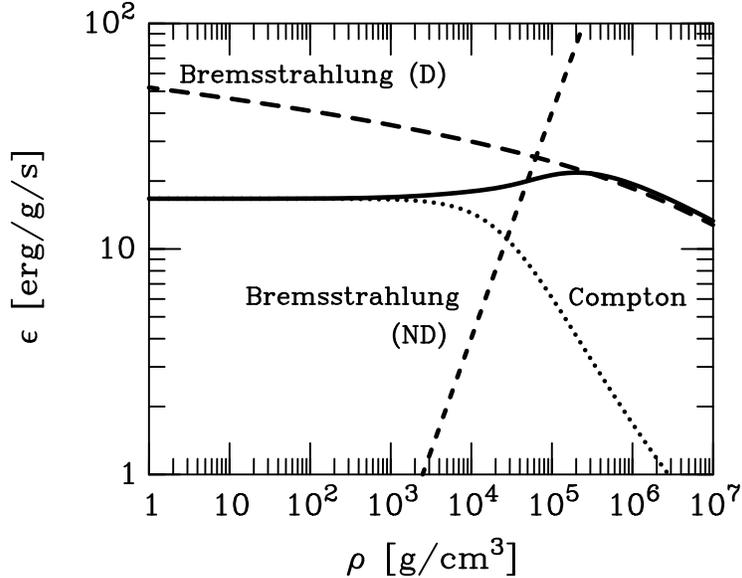

Figure 1: Axionic energy-loss rates for the processes discussed in the text for $\alpha_{26} = 1$, $T = 10^8$ K, and a composition of pure helium. The solid line is our interpolated emission rate.

this is a coincidence at the given temperature because the Compton and degenerate bremsstrahlung rates vary with different powers of $T$.

Dearborn, Schramm and Steigman [6] gave a table of their energy-loss rates. For a coupling constant $\alpha_{26} = 1$ and $T = 10^8$ K they used 20, 50, and 201 erg g$^{-1}$ s$^{-1}$ at densities $10^2$, $10^4$, and $10^6$ g/cm$^3$. At the highest relevant density this is about a factor of 10 above our rate.

## 3 Red Giant Evolution

In order to test the impact of axion emission on the evolution of red giants we have included the interpolation formula described in the previous section in our stellar evolution code in analogy to our previous study of non-standard neutrino losses [13]. We have then calculated several evolutionary sequences from the main sequence to helium ignition with different axion coupling strengths $\alpha_{26}$. We used a chemical composition corresponding to Mixture I of Ref. [13], i.e., to $Z = 10^{-3}$ and $Y_0 = 0.239$. The opacities were chosen for a Ross-Aller mixture; the impact of axion emission on the core mass is found to be the same for older Los Alamos ("AOL") [14] as well as the latest Livermore ("OPAL") [15] opacities, which have greatly improved the agreement between observations and stellar evolution theory in general. The mixing length parameter is taken to be 1.55. The plasma neutrino energy-loss rate was taken from Ref. [16]. The total stellar mass was $0.8\,\mathcal{M}_\odot$; mass loss on the red giant branch was ignored. For other aspects of our stellar evolution calculations see Ref. [13] and references therein.

We find that helium ignites at a core mass $\mathcal{M}_{\rm ig}$ which is increased by the $\alpha_{26}$-dependent amount which is given in Tab. 1 and shown in Fig. 2. The coupling strength $\alpha_{26} = 2$ corresponds approximately to the case where helium ignition was suppressed in the calculations of Dearborn, Schramm and Steigman [6] if one corrects for the overestimate of their emission rate. Even for stronger couplings helium still ignites in our calculations, although for our largest value ($\alpha_{26} = 8$) the core-mass increase is so large that, had we included mass loss, the entire envelope could have been consumed before helium had a chance to ignite.

Even though our calculations do not reproduce the suppression of helium ignition, which is an overly conservative criterion to constrain axion emission, we believe that the core mass *increase* alone can be used to derive a significant limit on $\alpha_{26}$.

Table 1: Increase of the core mass at helium ignition.

| $\alpha_{26}$ | $\delta\mathcal{M}_{\rm ig}\,[\mathcal{M}_\odot]$ |
|---|---|
| 0.0 | 0.000 |
| 0.5 | 0.022 |
| 1.0 | 0.036 |
| 2.0 | 0.056 |
| 4.0 | 0.080 |
| 8.0 | 0.111 |

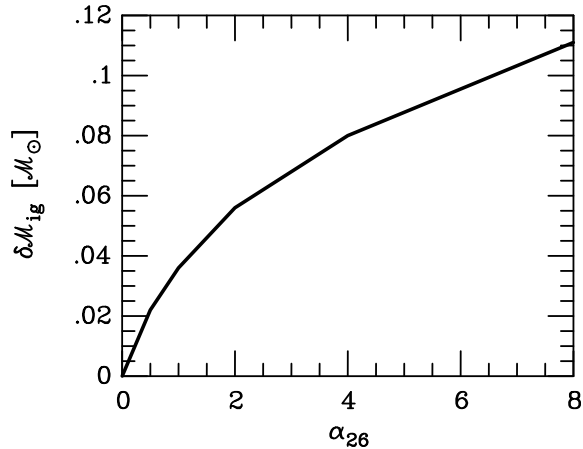

Figure 2: Increase of the core mass of a red giant at helium ignition due to axion emission.

## 4  Discussion and Summary

It was previously shown [17] that observations of globular cluster stars and of field RR Lyrae stars confirm the standard core mass at helium ignition $\mathcal{M}_{\rm ig}$ to within about 5%, i.e., to within about $0.025\,\mathcal{M}_\odot$. The main observational constraint is the maximum brightness reached by red giants, and the observed brightness of field RR Lyrae stars. We have previously used this method to constrain neutrino magnetic dipole moments [13].

$\mathcal{M}_{\rm ig}$ depends slightly on the total stellar mass and on the chemical composition (see [13, 17] for approximate analytic formulae); it is about $0.490 M_\odot$ for a helium content of 0.26 and a metallicity of 0.001. The systematic uncertainties of $\mathcal{M}_{\rm ig}$ due to possible deviations of the opacities from a Ross-Aller metallicity mixture, due to the standard mixing length treatment of convection, mass loss on the red giant branch, and the numerical shell-shifting technique all seem to be much smaller than this limit [13, 18, 19].

A core-mass increase of $0.025\,\mathcal{M}_\odot$ corresponds approximately to $\alpha_{26} = 0.5$, i.e., we find that globular cluster stars require that

$$\alpha_a \lesssim 0.5 \times 10^{-26} \qquad \text{or} \qquad m_a \lesssim 9\,\text{meV}/\cos^2\beta. \tag{14}$$

This is the strongest bound currently available on the axion-electron coupling,

but it is not in conflict with the interpretation that axions could contribute significantly to the cooling of ZZ Ceti stars such as G117–B15A.